\documentclass[prd,aps,onecolumn,floatfix,showpacs,preprintnumbers,amsmath,amssymb]{revtex4}
\usepackage{amsmath}
\usepackage{graphicx}
\usepackage{dcolumn}
\usepackage{bm}
\usepackage{amssymb}
\usepackage{latexsym}

\def\be{\begin{equation}}
\def\ee{\end{equation}}
\def\ba{\begin{eqnarray}}
\def\ea{\end{eqnarray}}

\usepackage{color}

\begin{document}


\title{Disruption of Cosmic String Wakes by Gaussian Fluctuations}

\author{Disrael Camargo Neves da Cunha \footnote{Email: camargod@physics.mcgill.ca}}
\affiliation{Department of Physics, McGill University, Montr\'eal, QC, H3A 2T8, Canada}

\author{Robert H. Brandenberger\footnote{Email: rhb@physics.mcgill.ca}}
\affiliation{Department of Physics, McGill University, Montr\'eal, QC, H3A 2T8, Canada}

\author{Oscar F. Hern\'andez\footnote{Email: oscarh@physics.mcgill.ca}}
\affiliation{Department of Physics, McGill University, Montr\'eal, QC, H3A 2T8, and
Marianopolis College, 4873 Westmount Ave., Westmount, QC H3Y 1X9, Canada}


\begin{abstract}

We study the stability of cosmic string wakes against
the disruption by the dominant Gaussian fluctuations
which are present in cosmological models. We find
that for a string tension given by $G \mu = 10^{-7}$
wakes remain locally stable until a redshift of $z = 6$,
and for a value of $G \mu = 10^{-14}$ they are stable
beyond a redshift of $z = 20$. We study a global stability
criterion which shows that wakes created by strings
at times after $t_{eq}$ are identifiable up to the present
time, independent of the value of $G \mu$. Taking into
account our criteria it is possible to develop
strategies to search for the
distinctive position space signals in cosmological
maps which are induced by wakes.

\end{abstract}

\maketitle

\section{Introduction}

Cosmic strings exist as solutions of the field equations in many
particle physics models beyond the Standard Model. A sufficient
criterion is that the vacuum manifold ${\cal M}$ of the model (the
space of field configurations which minimize the potential energy
density) has non-vanishing first homotopy group $\Pi_1({\cal M}) \neq $1.
Roughly speaking the condition is that the vacuum manifold has the
topology of a circle. A simple causality argument \cite{Kibble} leads
to the important conclusion that in models which admit cosmic
string solutions, a network of such strings inevitably forms during
the symmetry breaking phase transition in the early universe
and survives to the present time (see \cite{CSrevs} for reviews
of the role of cosmic strings in cosmology). Cosmic strings
carry energy and hence induce gravitational effects which can
lead to signatures in cosmological observations. The strength
of these effects is proportional to the string tension $\mu$
which in turn is given (up to a numerical constant) by $\eta^2$,
where $\eta$ is the scale of symmetry breaking at which the
strings are formed. Hence, searching for cosmic strings in
cosmological observations is a way to probe particle physics
beyond the {\it Standard Model} which is complementary to
accelerator searches (which can only probe new physics at
low energy scales) \footnote{See \cite{RHBCSrev} for an elaboration
on this theme.}.

Based on analytical arguments \cite{CSrevs} it is expected
that the distribution of cosmic strings will take on a ``scaling solution''
according to which the statistical properties of the distribution
of strings are independent of time if all lengths are scaled to the
Hubble radius $H^{-1}(t)$ (where $H(t)$ is the cosmic expansion
rate at time $t$). The distribution of strings consists of a network
of infinite strings with mean curvature radius and separation
$c_1 t$ (where $c_1$ is a constant of order one whose precise value 
needs to be determined in numerical simulations) \footnote{We are
here considering a simplified ``one-scale-model'' of strings.}
and a set of string loops which are the remnants of intersections
of long string segments. Numerical simulations \cite{CSnum}
have confirmed that the distribution of strings takes on a
scaling solution.

String loops oscillate and gradually decay by emitting gravitational
radiation. Long string segments moving through the plasma of the
early universe will lead to nonlinear overdensities in the plane
behind the moving string. These are called string {\it wakes} \cite{wake}.
Wakes are formed because the geometry of space perpendicular to a
long string segment is conical with deficit angle
\be
\alpha \, = \, 8 \pi G \mu \, ,
\ee
where $G$ is Newton's gravitational constant \cite{CSgrav}. A
string moving through the plasma with a velocity $v$ perpendicular
to the tangent vector of the string will lead to a velocity perturbation
\be
\delta v \, = \, 4 \pi G \mu v \gamma(v) 
\ee
from both sides towards the plane behind the moving string (where
$\gamma(v)$ is the relativistic gamma factor associated with the
velocity $v$). In turn, this leads to a thin region behind the
string with twice the background density, the {\it wake}. The
dimensions of the wake behind a string at time $t_i$ are
\be \label{dims}
c_1 t_i \, \times \, v \gamma(v) t_i \, \times \, 4 \pi G \mu v \gamma(v) t_i \, ,
\ee
where the dimensions are length along the string, depth of the wake in
direction of string motion, and mean thickness of the wake, respectively. We will denote these dimensions by $\psi_{1}$, $\psi_{2}$ and $\psi_{3}$ respectively, when using comoving coordinates.

Cosmic string loops accrete matter in a rougly spherical way and
give rise to density fluctuations which are hard to tell apart from
fluctuations formed by other point sources. String wakes, on the other
hand, give rise to signals with a clear geometrical signature, and
have hence been the focus of a lot of recent work (see e.g. \cite{RHBCSrev2}).
Long cosmic string segments produce line discontinuities in CMB
(cosmic microwave background) temperature maps \cite{KS}. The
contribution to the power spectrum of cosmological perturbations is
scale-invariant \cite{CSearly}. However, the fluctuations are active
and incoherent \cite{active} and hence do not lead to acoustic oscillations
in the angular power spectrum of CMB anisotropies. At the present
time, the angular CMB power spectrum in fact provides the most
robust upper bounds on the string tension \cite{CSbound} 
(see the introduction of~\cite{WFpaper} for a more detailed discussion on string tension limits as well as \cite{others} for earlier studies)
\be
G \mu \, < \, 1.3 \times 10^{-7} \, .
\ee
Hence, it follows that cosmic strings are only a sub-dominant 
component to the power spectrum of perturbations. The
dominant contribution must be due to almost Gaussian
and almost adiabatic fluctuations such as those produced
by inflation (or by alternatives to inflation such as {\it String Gas Cosmology}
\cite{SGC} or the {\it Matter Bounce} \cite{MB}).

Whereas overall cosmic strings are a sub-dominant component to
structure formation, string wakes can nevertheless give rise to prominent
signatures in position space maps. They give rise to a network of
edges in CMB temperature maps across which the temperature
jumps \cite{KS}, rectangles in the sky with a specific CMB polarization
signal (statistically equal E-mode and B-mode polarization with a
polarization angle which is uniform over the rectangle and whose
amplitude has a linear gradient \cite{Holder1}), and thin wedges of
extra absorption or emission in 21cm redshift maps \cite{Holder2}
(see also \cite{WFpaper, follow}).
These features are most prominent at high redshifts when string
wakes are already nonlinear fluctuations but the Gaussian fluctuations
are still in their linear regime. The cosmic string signals are also
most easily visible in position space maps (e.g. with edge detection
algorithms \cite{Danos}), whereas the distincitve stringy features
are washed out in power spectra (see e.g. \cite{Salton}).

At early times, cosmic strings dominate the nonlinearities in the
universe, the reason being that wakes are nonlinear perturbations
beginning at the time they are formed, whereas Gaussian
perturbations are linear at early times. At late times, however, 
the Gaussian fluctuations dominate the
structure in the universe. Most of the nonlinearities at the
present time are due to the Gaussian fluctuations. The question
we wish to address in this paper is whether the string-induced
inhomogeneities, which at early times are clearly visible, 
are still observable as coherent objects in
position space maps at later times (in particular times after
reionization). Concretely, we wish to study whether
string wakes will remain coherent or whether they are
disrupted by the Gaussian fluctuations. This analysis is
a crucial preliminary step towards identifying string signals
at low redshifts, e.g. in 21cm redshift maps at redshifts
comparable and smaller than the redshift or reionization,
or in large-scale structure redshift surveys.

In this paper we study various stability criteria
for string wakes. We study the stability of a wake
to local disruption and find the redshift above which a cosmic
string wake remains locally intact, as a function of $G \mu$.
However, even if Gaussian fluctuations cause the wake to
be locally disrupted, a global signal may remain. We
study a specific criterion which can be used to search
for the signals of primordial wakes. This analysis
shows that signals of string wakes remain from a global
perspective to the present time. Interestingly, the signals
can be identified independently of the value of $G \mu$, and
do not depend on whether the wakes are shock-heated or
diffuse (see \cite{Oscar} for a discussion of the difference
between these two cases). Our various stability criteria will
be relevant for developing robust observational strategies
to search for string wakes.

In the following section we give a brief review of cosmic string
wakes. In Section 3 we present a local stability condition
based on a {\it displacement condition}. In Section 4 we consider 
a local density contrast consideration. The
resulting stability condition shows that wakes are locally
disrupted by the Gaussian perturbations at a redshift lower
than some critical redshift which depends on $G \mu$. In 
Section 5 we discuss a global stability condition which shows that wakes are 
visible up to the present time independent of the value of $G \mu$. 

\section{String Wake Review}

Consider a string segment at time $t_i$ moving with velocity $v$
in direction perpendicular to the string. This segment will produce
an overdense region with twice the background density behind it
whose dimensions are given by (\ref{dims}). Once formed, this
wake will be stretched in the planar directions by the expansion of
space, and it will grow in thickness by accreting matter from above
and below. This accretion can be studied using the Zel'dovich
approximation \cite{Zeld}. We will consider wakes produced at
times $t_i > t_{eq}$, where $t_{eq}$ is the time of equal matter
and radiation. Those produced earlier cannot grow until
$t_{eq}$ and they will be hence be smaller. 

The thickness of the 
wake at time $t > t_i$ is determined by computing the comoving
distance $q_{nl}(t)$ of a shell of matter which is starting to collapse 
(``turning around'') onto the wake, i.e. for which
\be
{\dot{h}}(q_{nl}(t), t) \, = \, 0 \, ,
\ee
where the physical height is given by
\be
h(q, t) \, = \, a(t) \bigl[ q - \psi(q, t) \bigr] \, ,
\ee
where $a(t)$ is the cosmological scale factor and
$\psi(q, t)$ is the comoving displacement induced by
the gravity of the wake. A standard calculation
(see e.g. \cite{Lean, Holder1}) yields
\be \label{turn}
q_{nl}(t, t_i) \, = \, (z(t) + 1)^{-1}\  \frac{24 \pi}{5} v \gamma(v) G \mu (z(t_i) + 1)^{1/2} t_0 \, ,
\ee
where $z(t)$ is the cosmological redshift at time $t$ and $t_0$ is the
present time. At the turnaround $\psi(q_{nl},t)=\frac{1}{2} q_{nl}$. After turnaround, 
the shell of baryonic matter virializes at a distance which is half of the turnaround radius,
whereas the dark matter remains extended \cite{Sorn}.
Hence, the physical height of the dark matter wake at time $t$ is
\be \label{thick}
h(t, t_i) \, =  \, (z(t) + 1)^{-1}\ q_{nl}(t, t_i)
\, .
\ee
This is also the displacement which a particle experiences due to the wake
if this particle ends up at the edge of the wake.
We also denote the wake thickness in comoving coordinates by
\be
\psi_{3}(z) \, = \, \frac{24\pi}{5}10^{-7}(G\mu)_{7} v 
\gamma (v) t_{0}\frac{\sqrt{1+z_{i}}}{(1+z)} \label{eq:psi} \, ,
\ee
where $(G \mu)_7$ is the value of $G \mu$ in units of $10^{-7}$. 

The result (\ref{thick}) shows that the thickest wakes are those produced at
the earliest times, namely $t_i = t_{eq}$. The thickness of a wake is
obviously proportional to $G \mu$, and its comoving size grows linearly
in the cosmological scale factor $a(t)$, as expected from linear cosmological
perturbation theory.

\section{Displacement Condition}

In this section we will obtain a stability condition which is based on displacements 
induced by primordial Gaussian fluctuations. For simplicity we will restrict the analysis of this section to the matter dominated period. In the next section we will extend the validity range to include dark energy in the evolution of the growth factor.
The wake plane (formed by the $\psi_{1}$ and $\psi_{2}$ lengths) can be subdivided 
into pieces of area ${(\psi_{3})}^{2}$, where $\psi_{3}$ is the thickness of the wake 
in comoving coordinates. We will compute the displacement (in a direction 
perpendicular to the wake plane) which is coherent on this scale. In order to do 
this we will integrate in time the fluctuation of the peculiar velocity field on the 
scale $\psi_{3}$. 

If $S_{\psi_3}$ is the induced physical displacement, then 
\be
S_{\psi_{3}}(t)  \, <  \, h(t,t_i) \label{eq:displCond}
\ee
is a local displacement condition for the stability of the wake. To compute $S_{\psi_{3}}$ consider the continuity equation 
\be
\dot{\delta}+\frac{1}{a} \vec{\nabla} \vec{v} \, = \, 0 \, , 
\ee
where $\delta$ is the relative matter density contrast and $\vec{v}$ is 
the physical peculiar velocity field. Choosing a Fourier mode parallel to $\vec{v}$ and taking the modulus of the Fourier transform of the 
above equation we 
obtain a relation between the amplitudes of the velocity and density 
contrast fields in momentum space:
\be
\mid v_{k}(z)\mid \, = \, \frac{f a H } {k}\mid \delta_{k}(z)\mid
\ee
where we used $\delta(z)=g(z)\delta(0)$, $g(z)=D(z)/D(0)$ and $D(z)$ is the growth factor \cite{1992ARA..30..499C}. For the matter dominated period, $g(z)=1.29/(1+z)$ and the function $f(z)=\frac{a}{D(z)}\frac{d D(z)}{d a}$ is approximately one.

The contribution to the standard deviation of the peculiar velocity field on a 
scale $L=\frac{2\pi}{k}$ at redshift $z$ is denoted by $\Delta_{v}(k,z)$ 
and from the above equation we obtain
\be
\Delta_{v}(k,z) \, = \, a H {(\frac{L}{2\pi})}\Delta (k,z) \, ,
\ee 
where 
\be 
\Delta(k,z) \, \equiv \, \sqrt{\frac{k^{3}}{2\pi^{2}}P(k,z)} \label{eq:deltaSq}
\ee
is the dimensionless contribution to the standard deviation of the matter density fluctuations
$\delta$ on a length scale corresponding to $k$, given the dimensional
power spectrum $P(k,z)$ at redshift $z$.
The induced physical displacement $S_{\psi_{3}}$ is given by 
\be
S_{L} (z) \, = \, a \int^{z}_{z_i} a^{-1}( t^\prime) \Delta _{v}(k,z(t^\prime)) dt^\prime  
\ee
evaluated at $k=k_3$ where $k_{3} =2 \pi/ \psi_{3}$ is the
wavenumber associated with the 
comoving thickness $\psi_{3} (z)$ of the wake. The integral will be dominated by the upper limit of integration, therefore 
\be
S_{\psi_{3}} \, = \, a\frac{\psi_{3}}{2\pi}\Delta (\psi_3(z),z) \, ,
\ee
and the displacement condition (\ref{eq:displCond}) becomes
\be
\Delta (k_3 (z),z) \, < \, 2\pi \, . \label{eq:Displcond2}
\ee

When the above equation holds, the coherent displacement in a region perpendicular to the wake plane will be smaller than the wake thickness $\psi_{3}$. 
 This  displacement condition agrees to within one order of magnitude with the local delta condition of the next section and gives a physical interpretation to it. The above condition is valid during the matter dominated period, but in the next section this restriction will be extended to include the dark energy period.

\section{Local Delta Condition}

Another criterium for the stability of a wake can be obtained by
demanding that the r.m.s. Gaussian mass fluctuations $\Delta$ 
on the scale $k_3(z)$ of the wake thickness is smaller than unity, i.e.
\be
\Delta(k_{3}(z),z) \, < \, 1  \, . \label{eq:RMScond}
\ee 
We call (\ref{eq:RMScond}) the ``Local Delta condition'', which is stronger than (\ref{eq:Displcond2}). If this condition is
satisfied then the wake is locally stable. This condition can be justified 
by noticing that the matter density contrast $\delta$ in a volume within 
the wake fluctuates around 1 inside the wake and around zero outside, 
so if the standard deviation $\sigma$ of $\delta$ is of order one the wake 
matter signal will be lost. 

The late time power spectrum is obtained by multiplying the primordial power
spectrum by the square of a {\it transfer function} $T$ which comes from the
non-trivial evolution of fluctuations on sub-Hubble scales. Specifically, for
scales which enter the Hubble radius before the redshift $z_{eq}$ of equal
matter and radiation the fluctuations in matter on sub-Hubble scales grow only
logarithmically since the universe is dominated by a smooth radiation fluid
at these times and on these scales.

The late time power spectrum for a model with Gaussian fluctuations with
fixed spectral index is obtained from (\cite{Dodelson:2003}, page 184)
\be
P(k,z) \, = \, 2\pi^{2}{\delta_{H}}^{2}\frac{k^{n}}{{H_{0}}^{n+3}} T^{2}(k) g^{2}(z) \label{eq:powSpec}
\ee
where we use the expression 
given by \cite{1992ARA..30..499C} in the growth factor, which now includes dark energy , $T(k)$ is the transfer function,
$n$ is the scalar spectral index, and
${\delta_{H}}$ is the amplitude of $\Delta$ evaluated for a Fourier mode that 
corresponds to the Hubble scale. We choose a normalization that gives 
$\sigma_8=0.83$, where $\sigma_8$ is the rms fluctuation smoothed 
on a scale $8\  {\rm Mpc/h}$ using a top-hat window function. 
We use $n=0.97$ and 
$\Omega_{\Lambda}=0.7$ . At this point, we will switch from
natural units to units used conventionally in cosmology, namely ${\rm Mpc}$ for lengths
and seconds for time. In these units $c=9.6 \times 10^{-15}\ {\rm Mpc/s}$, and 
the expression on the right hand side of (\ref{eq:powSpec}) has to be multiplied by 
$c^{n+3}$. We will also use $v\gamma (v) =c /{\sqrt{3}}$, $z_{i}=1000$ and 
$t_{0}=4.35 \times 10^{17} {\rm s}$

The transfer function $T$ from (\cite{Bardeen:1985tr} page 60) is used to 
obtain an analytic expression for $\Delta(k_{3})$, which together with the 
approximation $(k_{3}(z))^{-0.0145}\approx 1$ gives
 \be \label{result2}
 \Delta(k_{3}(z),z) \, = \, 0.607 \ln(1+22.7 k_{3}(z))g(z) \, .
 \ee
This computation of Delta can now be applied to either condition (\ref{eq:Displcond2}) or (\ref{eq:RMScond}). For example, using the ``Local delta condition'' (\ref{eq:RMScond}), we find that the disruption redshift, the redshift when
$ \Delta(k_{3}(z),z) \, = \, 1$ depends only logarithmically 
on the wake thickness and hence
on the value of $G \mu$. We see that wakes are
stable to fairly late times.

In Figure 1 we plot the value of $ \Delta(k_{3}(z),z) \, $ (vertical axis)
as a function of redshift (horizontal axis) for the value $(G \mu)_7 = 1$ 
(black line) and $(G \mu)_7 = 10^{-4}$ (gray line). The dashed horizontal
line is $\Delta = 1$. We see that the wake is locally stable for $z$ above 
approximately 6 in the case of $G\mu = 10^{-7}$ and for $z$ above 
approximately 11 in the case of $G\mu = 10^{-11}$.  

\begin{figure}
\includegraphics[height=8cm]{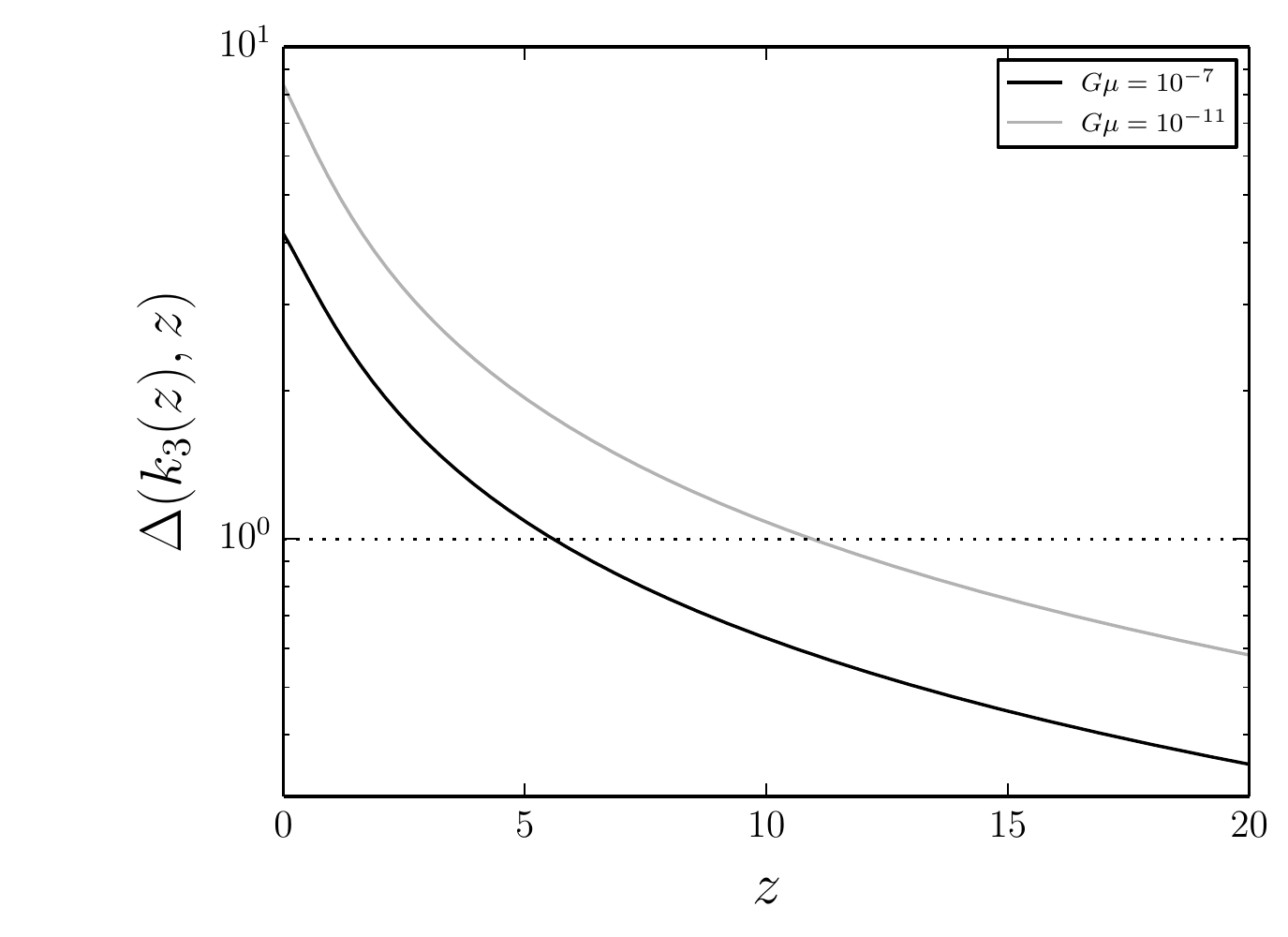}
\caption{Plot of $\Delta(k_{3}(z),z)$ (vertical axis) as a function of redshift 
$z$ (horizontal axis) for $G\mu = 10^{-7}$ (black line) and $G \mu = 10^{-11}$
(gray line).} 
\label{fig1}
\end{figure}

In Figure 2 we plot (the solid black line) the value of $(G \mu)_7$ (vertical axis) for which 
the stability condition of a wake ceases to be satisfied at redshift $z_d$ (horizontal axis). 
From this plot it follows that at $z_{d}=20$ all wakes $(G\mu) \geq 10^{-14}$ are stable. 
The dashed horizontal line is $(G\mu)_7 = 1 $, and we see that it intersects the solid black
line (which gives the value of $G\mu$ below which the wake is disrupted) at 
$z \approx 6$, confirming the result of Figure 1.
To obtain the value of $(G\mu)_{7}(z_{d})$ such that the wake will be disrupted at 
redshift $z_{d}$ (when the equality of (\ref{eq:RMScond}) is safisfied) 
we use 
\be
k_{3}(z_{d}) \, = \, 113(1+z_{d})/(G\mu)_{7}
\ee
in (\ref{result2}) to obtain
\be
  (G\mu)_{7}(z_{d}) \, = \, \frac{2565(1+z_{d})}{e^{1/0.607 g(z)}-1} \, .\label{eq:RedshiftDisruption}
\ee
  
\begin{figure}
\includegraphics[height=8cm]{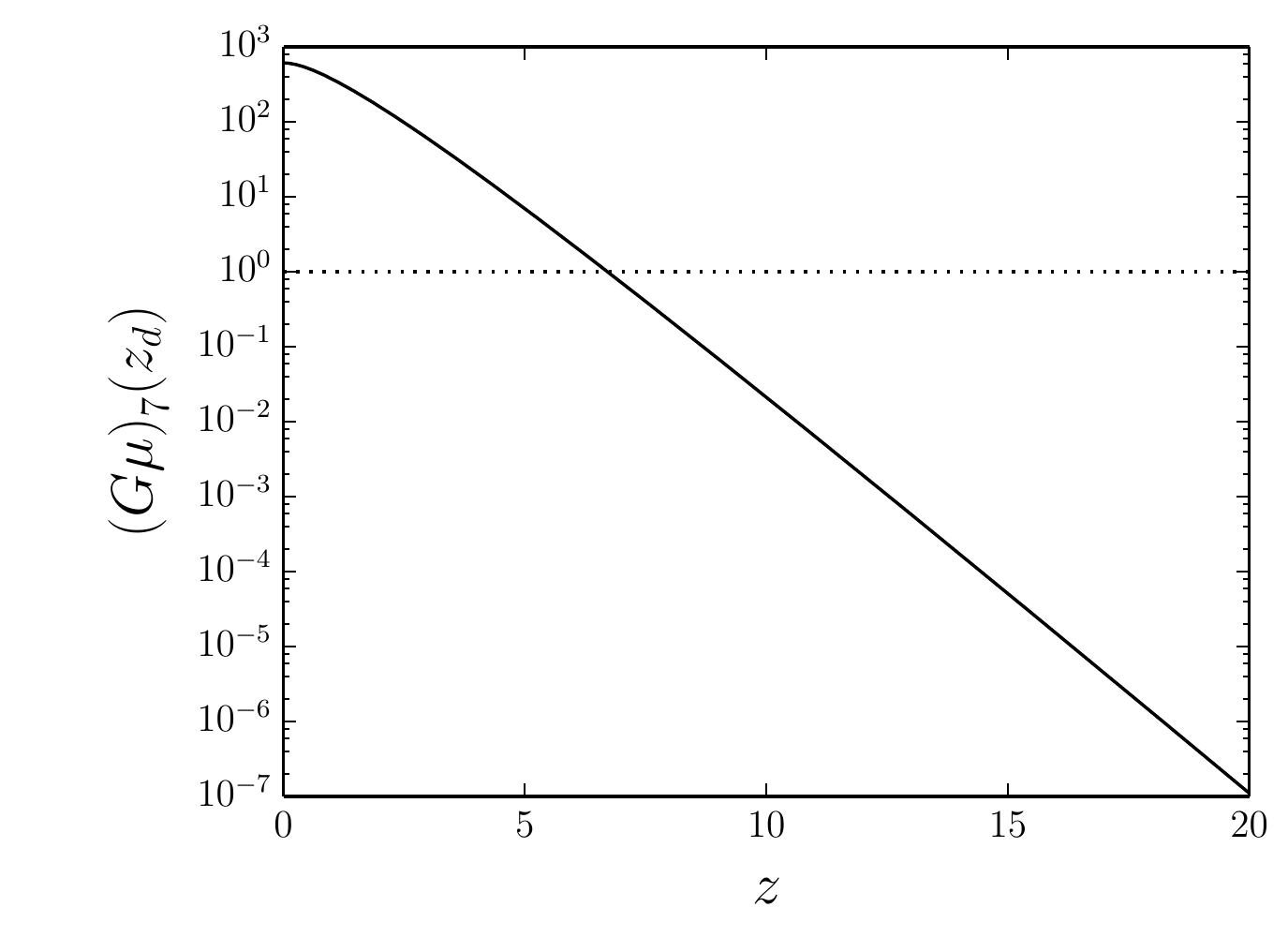}
\caption{The value of $G \mu$ (vertical axis in units of $10^{-7}$) above which the local Delta wake
stability condition is satisfied as a function of redshift $z$ (horizontal axis).} 
\label{fig3}
\end{figure}

\section{Global Sigma Condition}
  
The local Delta stability condition studied in the previous section is a very 
strict condition. It is
demanding that no section of the wake gets moved on a scale of the
wake thickness. A less restrictive condition is to demand that the wake
remains visible if we probe space with a filter which has the shape of the
three dimensional extended wake, i.e. which has two large dimensions
given by the length and depth of the wake, respectively, and one small
dimension given by the wake thickness. We call the resulting condition
the ``Global Sigma Condition''.

The variance of $\delta_{w}$ for a non-isotropic window function $\tilde{W}_{w}$ is given by
\be 
\sigma_{w}^{2} \, = \, 
\frac{g^{2}(z)}{{(2 \pi)}^{3}}\int_{-\infty}^{\infty} \int_{-\infty}^{\infty} \int_{-\infty}^{\infty} dk_{1} dk_{2} dk_{3} 
P(\| \vec{k} \|) {\tilde{W}}_{w}^{2}(\vec{k},z) \label{eq:variance}
\ee
where $g(z)$ is the growth factor and $P$ is the power spectrum at the
present time. Note that we are working in terms of comoving
momenta. The ``Global Sigma condition''
then is
\be \label{global0}
 \sigma_{w} \, < \, 1 \, 
\ee
when we consider a window function whose two large dimensions are given
by the planar size of the wake which is fixed in comoving coordinates.

The first guess would be to choose the small dimension to be given by
the wake thickness which is increasing in comoving coordinates. Before
making this choice, however, let us choose the thickness of the
window to be fixed in comoving coordinates, and present a rough 
analytical analysis. The integral (\ref{eq:variance}) is essentially cutoff by 
the radial planar size that corresponds to the comoving
momentum $k_r$, and the orthogonal size that correspond to $k_3$, with
$k_3 \gg k_r$. We then obtain
\ba \label{global1}
\sigma_{w}^{2} \, & \sim & \, \frac{g^2(z)}{2 \pi^2} \int_0^{k_3} dk_3 \int_0^{k_r} k_r 
P(\sqrt{k_r^2 + k_3^2}) \nonumber \\
& \sim & \, \frac{g^2(z)}{4 \pi^2} k_r^2 \int_0^{k_3} dk_3 P(k_3) \, .
\ea
For a roughly scale-invariant power spectrum of Gaussian fluctuations,
the final integral is dominated by scales which enter the Hubble radius
at around $t_{eq}$ where the power spectrum turns over (i.e. changes
from scaling as 
$k^{-3}$ 
for large values of $k$ to scaling as $k$
for small values). Let us denote this value of $k$ as $k_{to}$. Then
(\ref{global1}) yields
\be \label{global2}
\sigma_{w}^{2} \, \sim \, \frac{g^2(z)}{4 \pi^2} \bigl( \frac{k_r}{k_{to}} \bigr)^2
\Delta(k_{to})^2 \, ,
\ee
where $\Delta(k)^2$ is given by (\ref{eq:deltaSq}). Note that the result is
independent of $k_z$ as long as $k_z \gg k_{to}$.

Our result (\ref{global2}) lets us draw important conclusions. Most importantly,
the global delta criterium (\ref{global0}) is independent of the thickness
of the wake, and hence independent of the string tension $G \mu$. The equation
(\ref{global2}) also shows that wakes with larger planar extent, i.e. those
laid down later, are easier to identify that smaller wakes. The dependence
on $k_r$ is linear. This prediction can be used as a consistency check on
the numerical analysis. 

Another nice feature about our result is that it tells us that we
can choose a window function with a width greater than what we
expect the local displacements of the wake to be. 

We now turn to the quantitative evaluation of the condition. First,
the comoving planar dimensions of the wake can be read off from (\ref{dims}).
They are
\be
\psi_{1} \, = \, \frac{c_1 t_{0}}{\sqrt{1+z_{i}}} \, , \label{eq:psix} 
\ee
\be
\psi_{2} \, = \, \frac{v \gamma t_{0}}{\sqrt{1+z_{i}}} \, , \label{eq:psiy}
\ee
The wake thickness in comoving coordinates depends on $z$ and is given 
by equation (\ref{eq:psi})

Considering a wake region $V$ centred at the origin of coordinate space in the 
form of a parallelepiped of volume $V_{w} = \psi_{1}\times \psi_{2}\times \psi_{3}$ 
the wake window function in real space becomes
\be
W_{w}(X,Y,Z) \, = \, \left\{
                 \begin{array}{ll}
                     \frac{1}{V_{w}} & \mbox{if } (X,Y,Z) \in V \\
                     0               & \mbox{if } (X,Y,Z) \not\in  V
                 \end{array}    
            \right. \label{eq:WFR}
\ee 
and the Fourier transform of the above quantity is
\be
{\tilde{W}}_{w}(k_{1},k_{2},k_{3},z)=\frac{1}{V_{w}(z)}[\frac{2\sin(k_{1}\psi_{1}/2)}{k_{1}}][\frac{2\sin(k_{2}\psi_{2}/2)}{k_{2}}][\frac{2\sin(k_{3}\psi_{3}(z)/2)}{k_{3}}] \, . \label{eq:WFF}
\ee
The variance of $\delta_{w}$ is given by (\ref{eq:variance}).
Replacing (\ref{eq:WFF}) into (\ref{eq:variance}) results in
\be
\sigma_{w}^{2}(z) \, = \, \frac{1}{{(2 \pi)}^{3}}{(\frac{g(z)}{V_{w}(z)})}^{2}\int_{-\infty}^{\infty} \int_{-\infty}^{\infty} \int_{-\infty}^{\infty} dk_{1} dk_{2} dk_{3} P(\| \vec{k} \|) {([\frac{2\sin(k_{1}\psi_{1}/2)}{k_{1}}][\frac{2\sin(k_{2}\psi_{2}/2)}{k_{2}}][\frac{2\sin(k_{3}\psi_{3}(z)/2)}{k_{3}}])}^{2} \label{eq:variance1}
\ee
Note that in the integrand above
the $2\sin(k_{3}\psi_{3}(z)/2)/k_3$ term together with the $\psi_{3}$ term of 
$V_{w}$ approaches 1 as $G\mu \to 0$. Since for large $k_3$ the power spectrum also goes to zero, we can take this last term as 1 and hence for small $G\mu$ $\sigma_{w}$ is 
independent of $G\mu$. This confirms our expectation from equation (\ref{global2}). But this does not mean string wakes are visible at arbitrarily low string tension. A wake should not be disrupted in order for it to be seen. In this sense the global delta condition is a necessary but not a sufficient reason for detection. Though very low $G\mu$ wakes may not be disrupted, they are not necessarily detectable, since cosmic string wake signals are proportional to the string tension (see introduction of Ref. \cite{WFpaper} for a more detailed discussion of this point).
We explicitly verified the independence of $\sigma_{w}$ on $G\mu$ by 
evaluating the above integral numerically for several values from $G\mu=0$ to $10^{-7}$. It was assumed that $v \gamma (v)={c/\sqrt{3}}$ and 
$z_{i}=1000$ . We find that $\sigma_{w}(0)=0.32$.  Note that the entire $z$ dependence for  $\sigma_{w}(z)$ is given by the $g(z)$ factor in front of the integral. Until the time when dark energy becomes important we have $g(z) \propto 1/(z+1)$ and 
\be
\sigma_{w}(z) = 0.32 g(z) \  . 
\ee
The plot of $\sigma_{w}(z)$ is shown in Figure 3.
We conclude that even if
the wake is locally disrupted, the overall density pattern remains manifest. Good strategies for
cosmic string searches need to take this result into account.

\begin{figure}
\includegraphics[height=8cm]{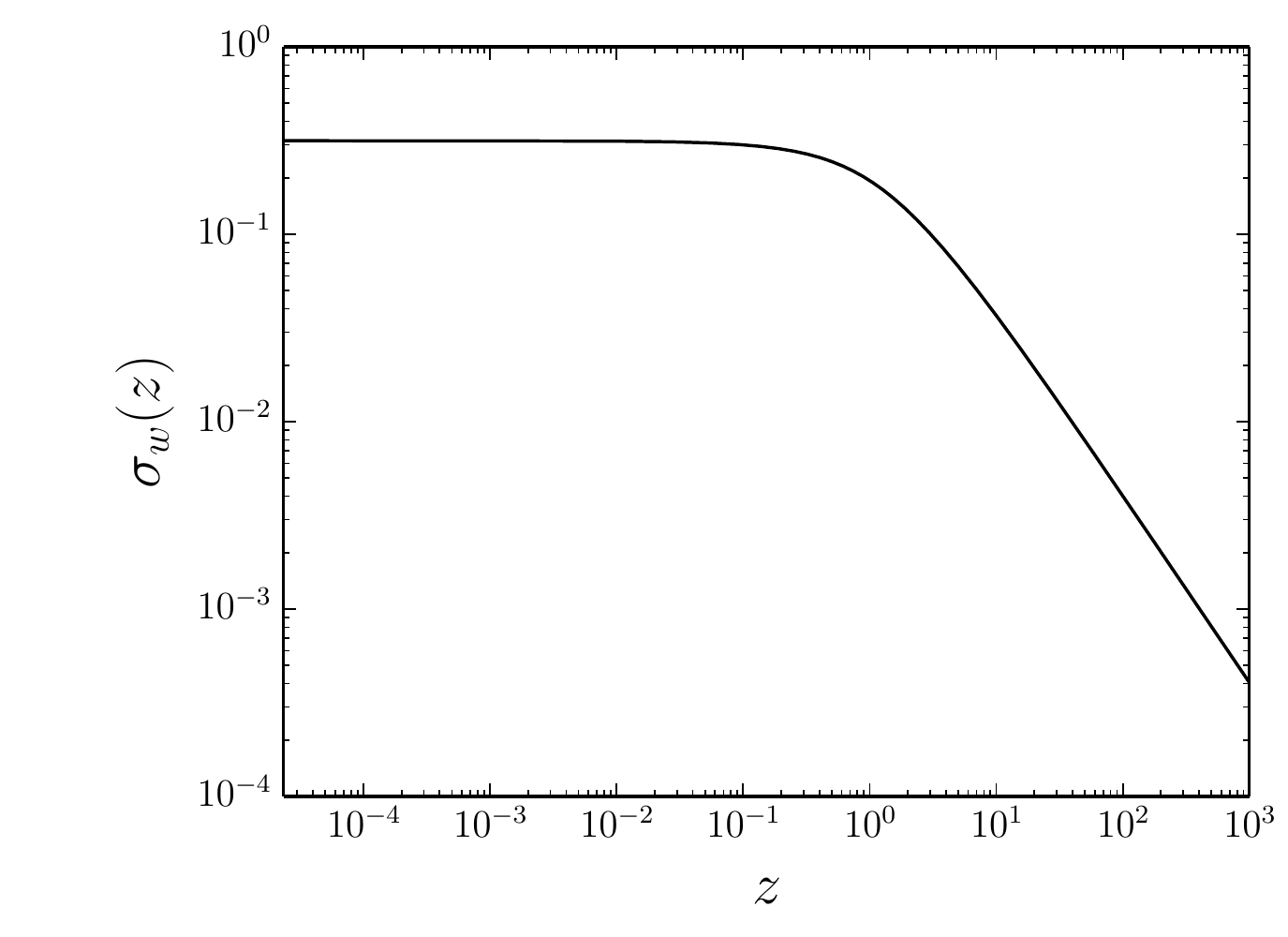}
\caption{The r.m.s. value of the density contrast of the Gaussian perturbations
in an anisotropic region which corresponds to the size of a wake produced
at $t_{eq}$ (vertical axis) as a function of redshift (horizontal axis). Note that the
density fluctuations remain smaller than one.} 
\label{fig4}
\end{figure}

Note that the $\sigma < 1$ conditions (both the global and local ones) are
good provided the fraction of additional matter (due to the wake) that is within a
region of the window function is of order one. In this case $\delta$ will fluctuate
around $1$ inside the wake and around $0$ outside, so $\sigma < 1$ will be a
good conditionl to distinguish 
between the presence and absence of a wake in a given region of space. 

\section{Discussion and Conclusions}

We have studied the disruption of a cosmic string wake by the gravitational
effects of the Gaussian fluctuations which dominate the current spectrum
of cosmological perturbations. At large redshifts the wakes are stable whereas
at smaller redshifts they are locally disrupted. The crossover redshift depends
on the string tension $G \mu$. For $G \mu = 10^{-9}$ the crossover redshift
is $z \simeq 11$. At redshifts greater than $z = 20$, wakes are stable down to
tensions of $G \mu = 10^{-14}$. To arrive at this result we investigated
both a local density contrast criterion and a displacement criterion. 

As an example, let us evaluate the possibility of seeing a $G\mu=10^{-9}$ cosmic string wake in a particular slice of the 21 cm maps from the SKA. Just above its local disruption of $z \simeq 11$, such a generically oriented wake has a  projected wake thickness $\Delta z_{\rm{wake}}$, 2 orders of magnitude smaller than the SKA redshift resolution of $\Delta z_{\rm SKA}=10^{-4}$. The planar size of the wake is $N=10^5$ times greater than the SKA angular resolution of $10^{-7}\ \rm{radians}$. Since the wake is not disrupted there is a slight overdensity over the entire $ 0.01\ \rm{rad} \times\ 0.01\ \rm{rad}$ region in redshift space. Consider those $N^2$ pixels that contain the wake as $N^2$ measurements in a no-wake theory. Knowing that the wake is undisrupted allows us to calculate the $\chi^{2}$ between a no wake theory and a theory with a wake for these pixels \cite{Agashe:2014kda}. We find that $\chi^{2}=N^{2}\times ({\Delta z_{\rm{wake}} / \Delta z_{\rm SKA}})^{2} = 10^{6}$ . Such a large $\chi^{2}$ results because we have assumed that all our pixels contain the wake. Obviously we have not addressed how to choose such candidate pixels, however here we wish only to show that a wake is visible in the scenario where our pixels do contain a wake.

The physical difference between the ``Local delta condition'' and the ``Global sigma condition'' is due to
the fact that in the local criteria, the relevant scale of the problem
is the wake thickness, and this scale is proportional to the string tension. 
On the other hand, as discussed above, the relevant scale for the global criteria is the
planar dimension of the wake which is independent of the string tension.

Even if a string wake is locally disrupted by Gaussian fluctuations,
it could possibly be identified using a ``Global Sigma Condition''. We have
computed the r.m.s. density contrast due to the Gaussian
fluctuations for an anisotropic window function whose
planar dimensions correspond to those of a wake, and whose thickness
is much smaller than the scale where the density power spectrum turns
over, and shown that the result is smaller than $1$ for all redshifts.
Hence, if we smooth the density field with such a window function,
then the wake will be visible even if it is locally disrupted.
This global condition is independent of the value of $G \mu$. We are looking 
for the dark matter component, so we do not have to consider (baryonic) 
diffuse wake corrections to the wake thickness.

Our work has implications for search strategies to find string
signals. Local features of wakes (e.g. discontinuity lines in
CMB polarization maps or sharp edges in three dimensional 21cm redshift
surveys) will only be visible for redshifts higher than
the crossover redshift determined by our local criteria. In contrast,
searches for string signals using global signals (e.g. statistical
analyses of maps obtained by smearing the original maps by an anisotropic window function of the shape of the expected
wake signal) will be promising even at very low redshifts.
We are currently studying this question.
 
\acknowledgements{We thank Yuki Omori for helpful discussions.
The work at McGill has been supported by an NSERC Discovery Grant, 
and by funds from the Canada Research Chair program. RB and OH have
also benefitted from support by a FQRNT group grant. OH was supported by the FQRNT Programme de recherche pour les enseignants de coll\`ege. 
DC thanks CAPES (Science Without Borders) for financial support.}

\end{document}